\documentstyle[12pt,epsfig,amsmath]{article}
\newcommand {\be}{\begin{equation}}
\newcommand {\ee}{\end{equation}}
\newcommand {\ba}{\begin{eqnarray}}
\newcommand {\ea}{\end{eqnarray}}
\begin{document}

\begin{flushright}
 \today
\end{flushright}

\begin{center}
{\Large \bf Coherence and  oscillations of cosmic neutrinos}\\
\vskip 0.5cm

{Yasaman Farzan$^{a}$~\footnote{yasaman@theory.ipm.ac.ir} and Alexei
Yu. Smirnov$^{b,c}$~\footnote{smirnov@ictp.it} }
\\

\vskip 0.2cm

{\it $^{a}$ Institute for research in fundamental
sciences
(IPM), P.O. Box 19395-5531, Tehran, Iran\\
$^b$ International Centre for Theoretical Physics, Strada Costiera
11,
34014 Trieste, Italy, \\
$^c$ Institute for Nuclear Research, Russian Academy of Sciences,
Moscow,
Russia\\
 }

\end{center}
\begin{abstract}
{ For cosmic neutrinos we study the conditions and the effects
 of  the coherence loss as well as coherent broadening of the spectrum.
 We evaluate the width of the neutrino wavepacket
produced by charged particles under various circumstances: in an
interaction-free environment, in a radiation-dominated medium
(typical of the sources of the gamma ray bursts) and in the
presence of a magnetic field. The effect of the magnetic field on
the wavepacket size appears to be more important than the
scattering. If the magnetic field at the source is larger than
$\sim$10 Gauss, the coherence of neutrinos will be lost while
traveling over cosmological distances. Various applications of
these results have been considered. We find that for large
magnetic fields ($B> 10^9$~Gauss) and high energies ($E_\nu>{\rm
PeV}$), ``coherent broadening"  can modify the energy spectrum of
neutrinos. In the coherent case, averaging out the oscillatory
terms of the probabilities does not induce any statistical
uncertainty beyond what expected in the absence of these terms. A
deviation from the standard quantum mechanics that preserves
average energy and unitarity cannot alter the picture. }

\end{abstract}

\section{Introduction}

Construction and operation of the neutrino telescopes open new
windows towards exploring the Universe. By studying the neutrinos
arriving at these telescopes from sources located at cosmological
distances, we can derive unprecedented information on the sources of
such neutrinos as well as on the  propagation and properties of the
neutrinos, themselves.

Operating detectors have not so far detected cosmic neutrinos. The
data accumulated during 2000-2004 by the AMANDA experiment shows
no indication of neutrino flux from beyond the atmosphere of the
Earth \cite{2000-2004} which   leads to an upper bound on the
point source flux \be E_\nu^2
\frac{d\Phi_\nu}{dE_\nu}\stackrel{<}{\sim} 10^{-8}~ {\rm GeV}~{\rm
cm}^{-2}~{\rm sec}^{-1}~{\rm sr}^{-1}\ , \label{amanda} \ee in the
energy range between 1.6~TeV and 2.5~PeV. This bound implies that
the total number of events per year at  a 1 km$^3$ scale detector
such as ICECUBE \cite{icecube-home} cannot be on average larger
than $10^4$. A host of theoretical models predict cosmic neutrino
fluxes in the (1 -100)~TeV energy range which  can be detected by
a 1 km$^3$ scale detector: {\it e.g.,} the fireball models for
Gamma Ray Burst (GRB) sources \cite{fireball,TeV}, models for
supernovae type I b/c \cite{Razzaque,Ando} and models for the
Active Galactic Nuclei (AGN) \cite{AGN}. The neutrino flux from
GRB in the (1-10)~TeV range can saturate the present bound from
AMANDA \cite{TeV}. We should prepare for such generosity of nature
and anticipate what can be learned about the neutrino properties
as well as the mechanism behind the production of neutrinos at the
source. Because of this prospect, we mainly concentrate on
neutrinos with energies in the  (1 - 100)~TeV range throughout the
present paper.

According to the models, neutrinos are produced through the
following chain of reactions \cite{Berezinsky:1970xj} \be p +
\gamma \to \Delta \to \pi^+ +X \ , \label{pigamma} \ee
\begin{eqnarray} \label{main-processes}\pi^+ &\to& \nu_\mu +\mu^+
\ ,\cr
 \mu^+ &\to& \nu_e+ \bar\nu_\mu +e^+ \ . \end{eqnarray} Along with the
process (\ref{pigamma}), both $\pi^+$ and $\pi^-$ can be produced
via the $p p$ collisions or via the collision of the protons with
the nuclei present in the medium. If the interaction rates of
$\pi$ and $\mu$ are negligible,  the flavor composition of the
neutrino flux at the source will be as the following \be
(F_e^0+F_{\bar{e}}^0):(F_\mu^0+F_{\bar\mu}^0):(F^0_\tau+F^0_{\bar\tau})=
1:2:0\ , \ee where $F^0_\alpha$ and $F^0_{\bar\alpha}$
respectively denote the fluxes of $\nu_\alpha$ and
$\bar{\nu}_\alpha$.

Neutrino oscillations alter the flavor composition of the neutrino
flux during propagation from the source to the detector. If the
flavor composition at the source is known, by studying the flavor
composition at the detector, one can extract  information on the
oscillation parameters \cite{Jezabek}. In particular in
\cite{Farzan:triangle,ISS}, as the key part of the program of
reconstructing the unitarity triangle and deriving the lepton sector
Jarlskog invariant, it was suggested to study the cosmic neutrinos
to extract the $U_{\mu 1}$ element of the
Pontecorvo-Maki-Nakagawa-Sakata matrix. There is also
rich literature on the possibility of directly determining the Dirac
CP-violating phase from the neutrino telescope data
\cite{CPcosmicCP}.

{ One of the important effects of neutrino propagation which can
in principle influence observations is  the loss of coherence.}
The different neutrino mass eigenstates having the same energy
have different velocities. Thus, the  wavepackets of the mass
eigenstates composing a neutrino state will come apart as they
propagate. If the traveled distance is so large that these
components completely separate from each other, they  will cease
to interfere at the detector.
In the  case of coherence loss,  the oscillatory terms of
the oscillation probabilities disappear.
Conversely, in the coherent
case, the different mass eigenstates   keep interfering and as a
result, the oscillatory terms are still present. Of course, even
in the latter case, because of the finite energy resolution of the
detector and the fact that the sources of different events are
located at various distances, the effects of the oscillatory terms
average to zero.  As discussed in \cite{Nussinov,Stod}, the two
cases are in practice indistinguishable. In this paper, we discuss
in detail under which conditions the coherence is lost. We also
evaluate the statistical uncertainty in the presence of the
oscillatory terms for the coherent case.

Coherent broadening of the wavepacket can lead to deformation of
the energy spectrum  which has direct observational consequences.
In this paper, we discuss under what circumstances, the effect
will be noticeable.

The paper is organized as follows.  In sec.~2, we elaborate on the
conditions of { the coherence loss}.
In sec.~\ref{loss}, we explore how the interactions of the parent
particles in the production region can affect the sizes of the
emitted wavepackets, and consequently, the loss of coherence.
In secs. \ref{lost-or-not-lost} and \ref{terrestrial}, we discuss
the applications of our finding
for neutrinos from various cosmic and terrestrial sources.
Sec.~\ref{broadening} is devoted to the discussion of coherent
deformation of the energy spectrum. In sec.~\ref{navasaan},  we
consider the effects of incomplete averaging. In
sec.~\ref{decoherence}, we { comment on} how a deviation from the
standard quantum mechanical evolution of states can affect the
results. A summary and the discussion of results are given in
sec.~\ref{conclusion}.

\section{Loss and restoration of coherence}
\subsection{Coherence loss \label{aver}}

The oscillatory terms from the oscillation probabilities can
disappear  in two physically distinct situations: (1) loss of
coherence; (2) averaging. { The first case is related to features
of production of neutrinos, whereas the second one  - of
detection.} { For the scope of further discussion we first
formulate the condition for loss of coherence in the momentum
space and then derive the same condition considering the problem
in the configuration space. The two approaches are equivalent and
give the same result.}

{ The phase of oscillation, $\phi$, equals}
 \be \phi(E, L) \equiv \frac{\Delta m^2
L}{2E}, \label{phase} \ee where $\Delta m^2 $ { is the mass
squared  difference} and $L$ is  the distance between the source
and the detector which hereafter will be referred to as the
baseline. Let us  define $\Delta E_L$ as energy difference for
which $\phi(E - \Delta E_L,L) - \phi(E,L) = 2\pi$. It is
straightforward to show that \be \Delta E_L \simeq \frac{4\pi
E^2}{\Delta m^2 L} = E \frac{l_\nu}{L}, \label{el} \ee where
$l_\nu = 4\pi E/\Delta m^2$ is the vacuum oscillation length.
Neutrinos traveling over cosmological distances are essentially on
their mass shells. We can therefore use the dispersion relation:
$E^2 = p^2 + m^2$. Introducing the uncertainty in the energy,
$\sigma_E$ and in the momentum \footnote{
 From now on, by $\sigma_p$ we implicitly mean the
width of the wavepacket in the longitudinal direction.},
$\sigma_p$, we find from this relation that $E \sigma_E = p
\sigma_p$. Width of the wavepacket in the energy-momentum space is
determined by $\sigma_E$ and $\sigma_p$. Furthermore, since
neutrinos are ultrarelativistic ($E\simeq p$), the uncertainty in
energy and momentum are practically equal: $\sigma_E \simeq
\sigma_p$. In what follows we will not distinguish between
$\sigma_E$ and $\sigma_p$. \footnote{{ Under certain
circumstances, like emission of neutrinos from crystals with
Mossbauer effect, the difference between the energy and momentum
uncertainties becomes important.}}

The interference between the effects of different mass eigenstates
disappears  if
\be
\label{11}\sigma_p > \Delta E_L=\frac{4\pi
E^2}{\Delta m^2 L} = E \frac{l_\nu}{L}.
\ee
(For an explicit
proof of the disappearance of the oscillatory terms see the
appendix.) Omitting the numerical factor, $4\pi$, we can rewrite
the condition in Eq.~(\ref{11})  as \be {\Delta m^2 L \over E}{\sigma_p \over E} \gg
1 . \label{loss-condition} \ee

Let us now reconsider the same problem in the configuration space.
The size of the wavepacket in the configuration space at the
production point is given by the inverse of the uncertainty
$\sigma_p$: \be \sigma_x \sim \frac{1}{\sigma_p}. \ee A neutrino
of a definite flavor  can be considered as a superposition of the
wavepackets of  different mass eigenstates. Due to the difference
between the group velocities of the mass eigenstates, \be \Delta v
= \frac{\Delta m^2}{2 E^2}, \label{difvel} \ee the corresponding
wavepackets  separate   in the course of propagation. The
separation, $d_L$, after propagating a distance  of $L$ is equal
to \be d_L = \frac{L}{v}\Delta v=L \frac{\Delta m^2}{2 E^2}.
\label{dl} \ee The separation in the configuration space is
complete when \be \sigma_x \ll d_L=L \frac{\Delta m^2}{2 E^2}.
\label{config} \ee At this point, the overlap of packets and
therefore interference effects disappear - the coherence is lost.
The condition in (\ref{config}) is equivalent to the condition in
(\ref{loss-condition}) in the energy-momentum space. { The
inequality (\ref{loss-condition}) as well as  (\ref{config}) is
called the condition for ``loss of coherence''. The coherence
length is defined as } \be L_{coh} = 4\pi \sigma_x
\frac{E^2}{\Delta m^2}. \label{cohl} \ee For $L>L_{coh}$, the
coherence is lost and if the coherence is not restored at the
detection (see subsec.~\ref{restoration} as well as the appendix),
the oscillatory terms disappear from the oscillation
probabilities.

For the typical source to detector distance  $L = 100$ Mpc, the
neutrino energy  $E = 10$ TeV and $\Delta m^2 = \Delta
m_{atm}^2=2.5 \cdot 10^{-3}$ eV$^2$, from Eq.~(\ref{config}) we
find { that the coherence is lost if the size of the wavepacket
satisfies} \be \sigma_x \ll d_L = 3 \cdot 10^{-3} {\rm cm}
\left(\frac{L}{100~{\rm Mpc}}\right) \left(\frac{\Delta m^2}{2.5
\cdot 10^{-3}{\rm eV}^2}\right) \left(\frac{10~ {\rm
TeV}}{E}\right)^2. \label{ev-sep} \ee Replacing $\Delta m^2$ with
$\Delta m_{sol}^2$, we obtain a value for $d_L$ that is about 30
times smaller. That is, if the size of the wavepacket is much
smaller than $ \sim 10^{-4}$ cm,  the oscillatory terms given by
$\Delta m_{sol}^2$  disappear.

In our consideration we have ignored  spread (widening) of each
wavepacket in the configuration space due to the presence of
different energies/momenta in the packet. For a given mass
eigenstate, the  group velocities of the components of the
wavepacket with $\Delta E = \sigma_E$ are different: $\Delta v =
(m^2/E^2)( \sigma_E/E)$. The spread of the wavepacket over a
distance $L$ is therefore equal to \be \sigma_{L} = |\Delta v L|=
\frac{m^2}{E^2}\frac{\sigma_E}{E} L=  \left(\frac{m}{E}\right)^3
\frac{1}{\sigma_x m}L. \ee Comparing this expression with the
separation  (\ref{dl}) we find  that the condition $d_L \gg
\sigma_{L}$ implies \be \frac{\sigma_E}{E} \ll \frac{\Delta
m^2}{m^2}, \label{as-long-as} \ee where $m$ is the heaviest
neutrino mass. The minimal value on the RHS  corresponds to the
degenerate neutrinos.  Eq. (\ref{as-long-as}) is fulfilled for all
situations under consideration in this paper. That is, the
separation of the wavepackets is more important than their spread.

\subsection{Restoration of coherence \label{restoration}}

Separation of the wavepackets does not mean yet  that
{ the effects of} oscillatory terms
are absent { in observation}.
If the time scale of the detection process is larger than the time interval
between the arrival of the successive wavepackets composing a
single state, the wavepackets still interfere at the detector
despite the { spatial} separation~\cite{Nussinov}. In other words, { if the
detector has large enough ``time memory''}
the detection restores the coherence. The coherence is determined
both by the production and detection. In this sense,  there is a
symmetry between the source and detector.

 Consider a neutrino wavepacket whose
components are completely separated: {\it i.e.,}
$\sigma_x\sim\sigma_p^{-1}\ll d_L$,  where $d_L$ is given in
Eq.~(\ref{dl}).   If the energy of the neutrino is measured with precision better than $d_L^{-1}$, the coherence will be restored \cite{Nussinov}.
 According to the energy-time
uncertainty principle, such a measurement takes a time longer than
$d_L/c$ during which the second component arrives at the detection
point. { However}, in the realistic situations under
consideration, restoration of the coherence at detection, which
requires determination of energy with precision  better than
$d_L^{-1} \sim 10^{-2}~ {\rm eV}(100~{\rm Mpc}/L)(2.5\times
10^{-3}~{\rm eV}^2/\Delta m^2)(E/10~{\rm TeV})^2$ (see
Eq.~(\ref{dl})),  does not take place. Suppose that a hypothetic
detector reaches such a sensitivity; then,  for each individual
neutrino state, coherence is  restored. However, even in this,
once we accumulate data from different sources located at various
distances, the oscillation pattern will disappear.
\section{Width of the neutrino wavepacket\label{loss}}

In this section, we evaluate the  width of a neutrino wavepacket
produced through the processes shown in Eq.~(\ref{main-processes})
in an environment such as the relativistic jets which, according to
the fireball models, are present  in the GRB sources. We discuss the
dependence of the wavepacket size on the density and the strength of
the magnetic field at the production region. We also discuss if a
neutrino produced in such an environment  loses its coherence
traveling over cosmological distances.

\subsection{Neutrino produced by decay of ``free''  pions
and muons \label{free}}

In this subsection, we  consider the  case in which the parent
charged particles (pions and muons) do not undergo any interaction
with matter or with the magnetic field before  decay. The
consideration applies for any environment
 in which the mean free path and the Larmor radius of the parent particle
are much larger than the distance that the charged particle travels
before decay.
 The cosmogenic neutrinos
 ({\it i.e.,}
 the neutrinos that are
produced through the interaction of the cosmic rays with the
galactic and intergalactic medium) can be considered as an example
for realization of such a situation.

 The width of the
wavepacket of the neutrino in the rest frame of the parent particle
can be estimated as \be \label{inverseTAU} \sigma_E^0 \approx
\sigma_p^0 \sim \frac{1}{\tau_0} \ , \ee where $\tau_0$ is the
lifetime of the parent particle in its rest frame ({\it i.e.,} for
the muon $\tau_0 =2.2\times 10^{-6}$~sec while for the pion $\tau_0
=2.6\times 10^{-8}$~sec).

 Suppose the pion propagates along the $z$-direction; {\it i.e.,} in the
observer frame,  the energy-momentum of the pion can be written as
$(E_\pi,0,0, p_\pi)$. Let us take the four-momentum of the
neutrino in the rest frame of the pion to be
$$(E^0_\nu,0,E^0_\nu\sin \xi,E_\nu^0\cos \xi),$$ where \be
\label{E0nu} E^0_\nu=\frac{m_\pi^2-m_\mu^2}{2m_\pi}\sim m_\pi/4
\ee and $\xi$ is the angle between neutrino momentum in the frame
of pion and the momentum of pion in the observer frame. The energy
of the neutrino in the observer frame can be then written as
$E_\nu=(E_\pi/m_\pi)E^0_\nu (1+v_\pi \cos \xi)$ where $v_\pi=
p_\pi/E_\pi\simeq 1$ is the velocity of the pion. Performing a
Lorentz transformation we find that in this frame \be
\label{sigmaEEE}\sigma_E = \frac{E_\pi}{m_\pi}(\sigma_E^0 + v_\pi
\sigma_p^0 \cos \xi )=\frac{E_\nu}{E^0_\nu}\sigma^0_E\sim
\frac{4E_\nu}{m_\pi}\frac{1}{\tau_0} \ .\ee Notice that
$\sigma_E/E_\nu$ is Lorentz invariant.

It is instructive to derive these results through alternative
approaches.  The length of the wavepacket in the rest frame of the
pion is equal to $c\tau_0$. It is straightforward  to show that in
the observer frame the wavepacket length is Lorentz contracted to
\be \label{sigmaXXX}\sigma_x\simeq{ \tau_0 \over \gamma (1+\cos
\xi)}\simeq \frac{E_\nu^0}{E_\nu}\tau_0 \sim
\frac{m_\pi}{4E_\nu}\tau_0\ .\ee
The last equation is for neutrinos emitted in the forward direction
$\xi = 0$. As expected,
Eqs.~(\ref{sigmaEEE},\ref{sigmaXXX}) yield $\sigma_x=\sigma_E^{-1}$.

We can rederive Eq.~(\ref{sigmaXXX}) by using the fact that the
number of wave periods  in the wavepacket is a Lorentz invariant:
$$
\frac{\sigma_x}{\lambda} = \sigma_x E_\nu = {\rm invariant}.
$$
In the rest frame of the pion this number is $\tau^0 p_\nu^0 =
\tau^0 E_\nu^0$, so \be \sigma_x = \tau^0 \frac{E^0_\nu}{E_\nu} \sim
\tau^0 \frac{m_\pi}{4E_\nu}. \label{sigma-x} \ee Neutrinos emitted
in the forward direction have higher frequency (energy) but shorter
packet (pulse).

Another Lorentz invariant that one can use to determine  $\sigma_x $
in the observer frame is the
number of the oscillation periods before the system loses its
coherence. This number is given by the ratio of the coherence length
$L_{coh}$ to the oscillation length, $l_\nu$. Using
Eq.~(\ref{cohl}), we find  \be \frac{L_{coh}}{l_\nu} = \sigma_x
E_\nu = {\rm invariant}, \ee which immediately leads to the same
result as (\ref{sigma-x}).

Numerically, for neutrinos produced by  pions we obtain \be \sigma_x
\sim 2 \times 10^{-3} {\rm cm}  \left(\frac{10~{\rm
TeV}}{E_\nu}\right). \ee For the neutrinos produced by  muons,
$\sigma_x$ is larger by the ratio of the lifetime of the muon to
that of the pion: $\sigma_x\sim 0.2~{\rm cm} (10~{\rm TeV}/E_\nu)$.

Let us consider the ratio of separation length and the length of
the wavepacket which, as discussed in sec. \ref{aver}, determines
the loss of coherence: \be \frac{d_L}{\sigma_x} = \frac{\Delta m^2
L}{2 E_\nu^2} \frac{1}{\sigma_x} \sim \frac{\Delta m^2 L}{E_\nu}
\frac{2}{m_\pi \tau^0}. \ee (For neutrinos from the muon decay one
should substitute $m_\pi$  by $m_\mu$ and insert the lifetime of
muon for $\tau_0$.) Notice that this ratio depends on the first
power of the neutrino energy. Numerically, we find \be
\label{no-scatter} \frac{d_L}{\sigma_x} \sim 0.1
\left(\frac{\Delta m^2}{8 \times 10^{-5}~{\rm eV}^2}\right)
\left(\frac{L}{100~{\rm Mpc}}\right) \left(\frac{10~{\rm
TeV}}{E_\nu}\right)  \left(\frac{3 \times 10^{-8}~{\rm
sec}}{\tau_0}\right) . \ee This ratio shows that despite the
extremely long baselines,  coherence is in general maintained.
This is always correct for neutrinos with $E_\nu \geq 1$ TeV
originating from the muon decay (even if we take $L$ as large as
the present size of the Universe). Only for relatively low energy
neutrinos, $E_\nu \sim$ TeV, from the pion decay, the coherence
can be lost.

\subsection{Interaction with particles in medium\label{scattering}}

If the parent particle (pion, muon) scatters off the particles in
the medium,  the length of the wavepacket of the produced neutrino
will be  shorter than in the interaction-free case. Here, for
illustrative purposes we consider a radiation dominated environment
as expected in  most models for the GRB sources. According to the
fireball models \cite{TeV,100TeV}, neutrinos at the GRB sources are
produced in relativistic jets which
have boost factors ($\Gamma_{jet}$)
of order of $ 10^2$ in the observer frame. In the frame comoving with the jet,
which we will call the jet-frame,  photons have a thermal distribution with
a typical temperature of $T_\gamma \sim$ few keV \cite{TeV}. In this
environment, the main interaction process for the charged particle
is scattering off the photons.

 Let us consider the
relevance of collisions and obtain the size of the neutrino
wavepacket $\sigma_x^{jet}$ in the jet-frame. The most energetic
pions (and muons) are supposed to move in the direction of jet and
therefore their energy in the jet frame can be estimated as \be
E_{\pi(\mu)}^{j} = \frac{E_{\pi(\mu)}}{\Gamma_{jet}} \approx
100~{\rm GeV} \left(\frac{E_{\pi(\mu)}}{10~{\rm TeV}}\right)
\left(\frac{10^2}{\Gamma_{jet}} \right), \ee where $E_\pi$
($E_\mu$) is the energy of the pion (the muon) in the observer
frame.

The average distance that the parent particle travels between two
successive collisions can be estimated as \be \label{mean-free-path}
\ell_{\rm col}\sim \frac{1}{\sigma n_\gamma}\ , \ee where
$n_\gamma={[2 \zeta(3)/\pi^2] T^3}$ is the photon density and
$\sigma$ is the total cross-section of the scattering. In the
photon-pion   center of mass system, the momentum of the pion  is of
order of $ \sim E_{\pi}^{j} T_\gamma /m_\pi \sim 4~{\rm MeV} \ll
m_\pi$; that is, in this frame the pion is non-relativistic. In
this regime, the cross-section of the Compton scattering  is given
by the Thompson formula:
\begin{equation}
\label{cross-compton} \sigma={8\pi \alpha^2\over 3m_\pi^2}.
\end{equation}
  In general, to calculate $\ell_{col}$, we
should  take into account the energy distribution of the
photons in the thermal bath but since in this non-relativistic regime
the cross-section is not sensitive to the momentum of the photon, the
approximation we have made  is justified. Inserting $n_\gamma$ and
$\sigma$ in Eq.~(\ref{mean-free-path}), we obtain
$$\ell_{\rm col}\sim 3 \times 10^{4}~{\rm cm} \left( {4~{\rm keV}
\over T_\gamma}\right)^3.$$
In the case of  muon scattering, $m_\pi$
has to be replaced by $m_\mu$. Notice that the mean  free path is
approximately the same for muons and pions.

In the jet-frame the pions are ultra-relativistic so the majority
of the produced neutrinos   are emitted within a narrow cone with
an opening angle of $m_{\pi}/E_{\pi}^j$, oriented along the
momentum of the parent particle. We will refer to this cone as
``the emission cone." The energy transfer in each collision is
negligible ({\it i.e.,} the collisions are elastic) and only the
momentum of the charged particle {  randomly changes direction} by
an angle of size $\sim T_\gamma/m_{\pi}$.  The smallness of the
momentum change in each collision  ({\it i.e.,}
$T_\gamma/m_{\pi}\ll m_{\pi}/E_{\pi}^j$) implies that if the
direction of the momentum of the parent particle before collision
is such that the line of sight lies within the emission cone,
after the collision, the line of sight will still remain within
the emission cone.

All the consideration we had so far in this sub-section holds
equally for the pion and the muon. However, we should bear in mind
that the neutrino production by the former takes place via a
two-body decay while from the latter it takes place via a three-body
decay. As a result, the widths of the neutrino wavepackets produced
by them are different. In the following we first discuss the pion
decay  and  we then consider the  muon decay.

 During the time between two successive collisions
 ($\ell_{col}/c$), the pion
 emits a wavepacket whose size  is of order of
 \be \label{pion-collision-width}\sigma_{x,\pi}^{jet}\sim \ell_{col}
 \frac{m_\pi}{E_\pi^j} \frac{E_\nu^0}{E_\nu^j}= \ell_{col}
 \frac{m_\pi}{E_\pi^j} \frac{m_\pi^2-m_\mu^2}{2m_\pi E_\nu^j}\ , \ee
where $E_\nu^0$ is the energy of the neutrino in the rest frame of
the pion (see Eq. (\ref{E0nu})). Notice that, in
Eq.~(\ref{pion-collision-width}), we have taken into account the
Lorentz contraction of the wavepacket as discussed in sec.
\ref{free} (see Eq.~(\ref{sigmaXXX})).

 After the collision, the momentum of the
pion is changed by $|\Delta \vec{p}_\pi^j| \sim T_\gamma
E_\pi^j/m_\pi$ in a direction transverse to the momentum ($\Delta
\vec{p}_\pi^j \cdot \vec{p}_\pi^j \ll |\Delta \vec{p}_\pi^j ||
\vec{p}_\pi^j|$). Writing the kinematics of scattering, we find that
the difference between the energy of the neutrino emitted in our
direction before and after the collision is $\sim T_\gamma
(E_\pi^j/m_\pi)^2$ which is much larger than
$(\sigma_{x,\pi}^{jet})^{-1}$ given in
Eq.~(\ref{pion-collision-width}). This means that the wavepackets
emitted before and after the collision are incoherent and the
effective size of the wavepacket in the jet frame is therefore given
by Eq.~(\ref{pion-collision-width}).

The spectrum of the neutrinos produced by the muon is continuous.
Thus, even after a collision the wavepacket of the neutrino will
have a component of the same energy { as before}. However, if we
wait long enough the direction of the parent particle will change
so much that the line of sight will exit the emission cone. After
this time, the amplitude of the neutrino emitted in the
observation direction is negligible. Thus, the size of the
wavepacket is determined by the time, $\Delta t$, during which the
line of sight  exits the emission cone. Let us estimate $\Delta
t$. As mentioned, after each collision the momentum of the parent
particle changes direction randomly  by an angle of size
$T_\gamma/m_\mu$ so, after $N$ collisions, the total rotation
angle on average amounts to $\sqrt{N} T_\gamma/m_\mu$. Thus, for
$E_\mu^j T_\gamma\ll m_\mu^2$,  after $N\sim (m_\mu^2/E_\mu^j
T_\gamma)^2$ collisions, the direction of the momentum changes by
an angle of $\sim m_\mu/E_\mu^j$ and the line of the sight
therefore exits the emission cone. Since the time between two
successive collisions is of order of $\ell_{col}/c$, we conclude
that
$$\Delta t\sim \frac{\ell_{col}}{c} \left({m_\mu^2\over E_\mu^j T_\gamma}\right)^2. $$
Taking into account the Lorentz contraction (see sec.~\ref{free})
for neutrinos emitted in the forward direction we obtain \be
\label{widthXMU}\sigma_{x,\mu}^{jet} \sim c\Delta t
\frac{m_\mu}{E_\mu^j}\frac{E^0}{E_\nu^j}\sim \frac{\ell_{col}}{3}
\frac{m_\mu^6}{T_\gamma^2(E_\nu^j)^2E_\mu^2},\ee where $E^0$ is
the energy of the neutrino in the rest frame of the muon which is
of order of $m_\mu/3$. {Notice that this relation is valid only if
$E_\mu^j T_\gamma\ll m_\mu^2$. For larger values of $E_\mu^j
T_\gamma$,  a single collision will be enough for the line of
sight to exit the emission cone ({\it i.e.,} $N=1$) and therefore
$\sigma_{x,\mu}^{jet}\sim \ell_{col}(m_\mu E^0)/(E_\mu^j
E_\nu^j)$.}

 As we discussed in the previous section, $\sigma_E/E$, or
 equivalently $E\sigma_x$, is invariant under the Lorentz
 transformations. Thus, from Eq.~(\ref{pion-collision-width}), we
 find that in the observer frame, the wavepackets of the neutrinos
produced by the pion   are of  size of \be \label{sigmaX-pi}
\sigma_{x,\pi}=\sigma_{x,\pi}^{jet} \frac{E_\nu^{jet}}{E_\nu}\sim
3\times 10^{-5} ~{\rm cm} \left(\frac{10~{\rm TeV}}{E_\nu}\right)^2
\left(\frac{4~{\rm keV}}{T_\gamma}\right)^3 \frac{\Gamma_{jet}}{100}
\ . \ee With this wavepacket size,  we obtain from
Eq.~(\ref{ev-sep}) that for neutrinos produced by the pions, the
wavepackets of all three neutrino mass eigenstates will be separated
on their way to Earth from the GRB sources.

Similarly, using the invariance of $E\sigma_x$ we obtain \be
\label{sigmaX-mu} \sigma_{x,\mu}=\sigma_{x,\mu}^{jet}
\frac{E_\nu^{jet}}{E_\nu}\sim 5 \times 10^{-3}~{\rm cm}
\left(\frac{4~{\rm keV}}{T_\gamma}\right)^5\left(\frac{10~{\rm
TeV}}{E_\nu}\right)^4 \left(\frac{\Gamma_{jet}}{100}\right)^3\ee
From Eq.~(\ref{ev-sep}) we conclude that unlike the case of
neutrinos from the pion decay, the different mass  components of
the neutrinos produced by the muon  will keep overlapping until
they reach the Earth. In other words, the coherence of the
neutrinos from the muon decay is maintained.

{In the energy-momentum space, the wavepacket sizes in
Eqs.~(\ref{sigmaX-pi},\ref{sigmaX-mu}) translate into \be
\frac{\sigma_{E,\pi}}{E_\nu}\sim 10^{-13}\left(
\frac{E_\nu}{10~{\rm TeV}}\right) \left( \frac{T_\gamma}{4~{\rm
keV}}\right)^3\left(\frac{100}{\Gamma_{jet}}\right) \ \ \ ({\rm
for} \ \ \frac{E_\nu T_\gamma}{\Gamma_{jet}} \ll m_\pi^2)
\label{sigmaE-pi} \ee and \be \frac{\sigma_{E,\mu}}{E_\nu}\sim
10^{-15}\left( \frac{E_\nu}{10~{\rm TeV}}\right)^3 \left(
\frac{T_\gamma}{4~{\rm
keV}}\right)^5\left(\frac{100}{\Gamma_{jet}}\right)^3  \ \ \ ({\rm
for}\ \ \frac{E_\nu T_\gamma}{\Gamma_{jet}} \ll m_\mu^2).
\label{sigmaE-mu} \ee
Increasing the energy, $\sigma_E$ increases. Notice however that
for $E_\nu T_\gamma/\Gamma_{jet}\stackrel{>}{\sim}m_\pi^2$, the
above formulas are not valid for two reasons: (i)  in this regime
the cross section is not given by the Thompson formula; (ii) as
discussed before, for high energies, Eq.~(\ref{widthXMU}) is not
valid. Taking into account these considerations we find that \be
\frac{\sigma_{E,\pi}}{E_\nu} \sim \frac{\sigma_{E,\mu}}{E_\nu}
\sim 5 \times 10^{-12}(\frac{T_\gamma}{4~{\rm keV}})^3  \ \ \ \
{\rm (for \ \ \frac{E_\nu
T_\gamma}{\Gamma_{jet}}\stackrel{>}{\sim} m_\pi^2)}.
\label{high-sigmaE}\ee
}
\subsection{The effects of  magnetic fields \label{magnetic}}

Let us discuss the effects of the  magnetic field  present in the
source on  the width of the neutrino wavepacket. The magnetic
field at the GRB sources can be as large as $\sim 10^9$~Gauss
\cite{TeV,Razzaque,Ando}. For simplicity, we assume that the
magnetic field is constant and uniform: ($\vec{B}=B \hat{z}$).
Later on, we will show that this assumption is justified. Notice
that although the magnetic fields in the sources under
consideration are quite large but still $e B\ll E^2$ and we can
therefore treat the effects of the magnetic field classically.
That is, we can describe the state of the charged particle as a
plane wave whose momentum
 in the direction transverse to the magnetic
field slowly rotates in time. As we will see, the decays of the pion
and  muon have to be treated differently.

 Let us  first discuss the pion decay. In
 the magnetic field, the charged pions move on circular (spiral)
trajectories with Larmor radius $R = E_\pi/eB$. The  energy of the
emitted neutrino in a two body decay depends on the angle between
its momentum and the momentum of the pion which, in turn,  is a
function of time. That is, the energy of the emitted neutrino at
time $t$, $E_\nu(t)$, differs from that at time $t +\Delta t$: \be
\Delta E_\nu\equiv E_\nu(t+\Delta t)-E_\nu(t). \label{condition}
\ee Taking into account the Lorentz contraction (\ref{sigmaXXX})
we find that the length of the neutrino wavepacket in the
configuration space emitted during the period $\Delta t$ from a
pion with energy of $E_\pi$ is given by
$$\sigma_x=(c\Delta t)
\frac{m_\pi}{E_\pi}\frac{E_\nu^0}{E_\nu},$$
where $E_\nu^0$ is the
energy of the neutrino in the rest frame of the pion ({\it see}
Eq.~(\ref{E0nu})).
 Consider segments of the pion trajectory for which $\Delta E_\nu$
defined in Eq.~(\ref{condition}) coincides with
$\sigma_E=\sigma_x^{-1}$: \be \label{segments} \sigma_E=\Delta
E_\nu=\frac{1}{\Delta t}\left(\frac{E_\pi}{m_\pi}\right)
\left(\frac{E_\nu}{E_\nu^0}\right)=\frac{1}{\Delta t}
\left(\frac{2 E_\pi E_\nu}{m_\pi^2-m_\mu^2}\right) . \ee The
wavepackets emitted during successive $\Delta t$ can be resolved
by a hypothetical detector with energy resolution $\sigma_E$ and {
time resolution} $\sigma_E^{-1}$, so their effects sum up
incoherently. Consequently, the coherence length of the neutrino
wavepacket is determined by $\sigma_E^{-1}$ which can, in turn, be
derived by solving Eq.~(\ref{segments}). For this, one needs to
determine $\Delta E_\nu$ or $E_\nu$ as a function of time.

The four-momentum of the pion  can be written as \be
\label{E-of-pion} (E_\pi, ~~p_\pi \sin \theta_\pi \cos\Phi(t) ,~~
p_\pi \sin \theta_\pi \sin \Phi(t) ,~~ p_\pi \cos \theta_\pi), \ee
where $\Phi(t)$ is the rotation phase: \be \Phi(t) = {eB (t -
t_0)\over E_\pi}\ . \ee  Notice that $E_\pi$ is constant.
Let us denote  the unit vector in the direction from the source to
the detector by $\hat{l}$, and define the axis $\hat{y}$ as $\hat{y}
= \hat{l}\times \hat{z}/|\hat{l}\times \hat{z}|$. In this frame, the
neutrino emitted at time $t$  towards the detector has the following
four-momentum \be \label{Enunu} E_\nu (t) (1,\sin \theta_\nu,0,\cos
\theta_\nu), \ee where $\cos \theta_\nu = (\hat{l} \cdot \hat{z})$
and \be \label{Enut} E_\nu(t)={m_\pi^2-m_\mu^2 \over 2[E_\pi - p_\pi
\cos (\theta_\nu-\theta_\pi)+2p_\pi \sin \theta_\pi \sin \theta_\nu
\sin^2 \frac{\Phi(t)}{2 } ]}. \ee Notice that due to the two-body
character of the pion decay, the neutrino energy  in the rest frame
of the pion is fixed, and in the observer frame it is uniquely
determined by the angle between the momenta of the pion and the
neutrino. From Eq.~(\ref{Enut}), we  obtain \be
\label{E-splitting}\Delta E_\nu = \frac{2 eB E_\nu^2\sin \theta_\nu
\sin \theta_\pi\sin\Phi}{m_\pi^2-m_\mu^2}\Delta t.\ee

Only a small fraction of the neutrinos (of order of
$(m_\pi/p_\pi)^2$) are emitted in directions for which the angle
between the neutrino and pion momenta is larger than $m_\pi/p_\pi$.
We therefore concentrate on the angles and emission time for which
the line of sight lies inside the emission cone: \be
\label{cone-condition}(\theta_\nu-\theta_\pi)\sim \sin \Phi(t)
\stackrel{<}{\sim} m_\pi/p_\pi \ll 1.\ee
 For such values of
$t$, $\theta_\pi$ and $\theta_\nu$, from
Eqs.~(\ref{segments},\ref{E-splitting}) we obtain \be
\label{width-g} \sigma_E \sim \left[ \frac{4  eB E_\nu^3 m_\pi\sin
\theta_\pi \sin \theta_\nu}{(m_\pi^2-m_\mu^2)^2}\right]^{1/2}  . \ee

Let us apply these results to neutrinos  which are produced in
relativistic jets \cite{TeV,100TeV}. We perform our analysis
in the (comoving) jet-frame where the electromagnetic field is
predominantly composed of the magnetic component \footnote{In the
observer frame, according to the Lorentz transformation in addition
to the magnetic field, an electric field will be also present.}.
 Using the invariance of $\sigma_E/E$ (see Eq.~(\ref{sigmaEEE})),
 we can immediately write
the  width in the observer frame, $\sigma_E$, in terms of the energy
width in the jet-frame, $\sigma_E^{jet}$: \be \frac{\sigma_E}{E_\nu}
= \frac{\sigma_E^{jet}}{E^{jet}} \sim \sqrt{\frac{eB E_\nu
m_\pi}{\Gamma_{jet} (m_\pi^2-m_\mu^2)^2}} \sim
 10^{-4} \left(\frac{100}{\Gamma_{jet}}~\frac{E_\nu}{10~{\rm
TeV}}~\frac{B}{10^7 ~{\rm Gauss}}\right)^{1/2},  \label{ratio-s}
\ee where we have taken  $4\sin \theta_\pi \sin \theta_\nu \sim 1$
and $E^{jet}_\nu \sim E_\nu/ \Gamma_{jet}$. The width of the
wavepacket in the configuration space  is given by \be
\label{sigmaX-B-pi}\sigma_x\sim\sigma_E^{-1}\sim 2\times
10^{-14}~{\rm cm} \left(\frac{\Gamma_{jet}}{100} \frac{10^7~{\rm
Gauss}}{B}\right)^{1/2}\left(\frac{10~{\rm
TeV}}{E_\nu}\right)^{3/2}.\ee Conferring the  estimations in
Eqs.~(\ref{sigmaX-B-pi}) and (\ref{sigmaX-pi}) we conclude that
for $B>10^{-11}$~Gauss, interaction of pions with the magnetic
field is more effective in shortening the wavepacket than their
scattering off the particles in the medium. That is, in practice,
the size of the wavepacket of neutrinos produced in a cosmic
neutrino source is given by Eq.~(\ref{sigmaX-B-pi}) rather than by
Eq.~(\ref{sigmaX-pi}).


Using the ratio in Eq.~(\ref{ratio-s}) and the condition shown in
Eq.~(\ref{loss-condition}), we find that  the loss of coherence
 over cosmological
distances occurs if \be \label{B-limit} B \gg 5 \times 10^{-13}
~{\rm Gauss} \left( \frac{8 \times 10^{-5}~{\rm eV}^2}{\Delta
m^2}\right)^2\left(\frac{100~{\rm
Mpc}}{L}\right)^2\frac{E_\nu}{10~{\rm
TeV}}\frac{\Gamma_{jet}}{100}.\ee

 Let us now consider neutrinos from
the muon decay. Due to  the three-body character of the decay, the
energy spectrum of the neutrino emitted in a given direction is
continuous. Thus, the consideration that we had in the case of the
two-body decay of the pion does not apply here.  Since the muon is
ultrarelativistic, the majority of the emitted neutrinos  are
oriented in the direction of the muon momentum within the emission
cone. Only a small fraction of $O(m_\mu^2/E_\mu^2)$ is emitted
outside the emission cone. Since the momentum of the muon rapidly
rotates, only for a short period, $\Delta t$, the line of
sight lies inside the emission cone.
A reasonable estimate for $\Delta t$ is the time interval during
which the angle between the line of sight and the momentum of the
muon is smaller than $m_\mu /E_\mu$: $(eB \Delta t /E_\mu)
\stackrel {<}{\sim} m_\mu /E_\mu$. We then find that in the
jet-frame \be \label{pulsed-neutrinos} \sigma_E^{jet}= (\Delta
t)^{-1} \left({E_\mu^{jet} \over m_\mu} \right)^2 =  {e B \over
m_\mu} \left(\frac{3E_\nu^{jet}}{m_\mu}\right)^2, \ee where we
have taken $E_\nu \sim E_\mu/3$. Using the invariance of
$\sigma_E/E$, we obtain \be
\label{muon-ratios}\frac{\sigma_E}{E_\nu}=\frac{\sigma_E^{jet}}{E_\nu^{jet}}=5
\times 10^{-8} \frac{B}{10^7~{\rm Gauss}}\frac{E_\nu}{10~{\rm
TeV}}\frac{100}{\Gamma_{jet}} \ , \ee where  $E_\nu$ is the energy
in the observer frame: $E_\nu=E_\nu^{jet}/\Gamma_{jet}$. Thus, in
the observer frame, the length of the wavepacket  is \be
\label{sigmaX-B-mu}\sigma_x\sim \sigma_E^{-1} \sim 5 \times
10^{-11}~{\rm cm} \left(\frac{10 ~{\rm TeV}}{E_\nu}\right)^2
\frac{10^7~{\rm Gauss}}{B}\frac{\Gamma_{jet}}{100} \ . \ee Again,
comparing the estimations in Eq.~(\ref{sigmaX-B-mu}) and in
Eq.~(\ref{sigmaX-mu}) we find that for  $B>10^{-1}$~Gauss, the
interaction of muons with the magnetic field is more effective in
shortening the wavepacket size  than their scattering off the
particles in the medium. In the presence of the magnetic field,
the wavepacket size is therefore given by Eq.~(\ref{sigmaX-B-mu})
rather than by Eq.~(\ref{sigmaX-mu}).

 To obtain complete loss of
coherence ({\it see,} Eq.~(\ref{loss-condition})), the following
condition has to be fulfilled: \be B\gg 10~{\rm Gauss}
\frac{100~{\rm Mpc}}{L} \frac{8 \times 10^{-5}~{\rm eV}^2}{\Delta
m^2} \frac{\Gamma_{jet}}{100}. \label{B-limit-muon}\ee

Notice that while in the case of the pion decay, $\sigma_E$ is
proportional to $B^{1/2}$ (see Eq.~(\ref{ratio-s})), in the muon
decay case, the  dependence on $B$ is linear (see
Eq.~(\ref{muon-ratios})). This explains the huge difference
between the lower limits on $B$ shown in
Eqs.~(\ref{B-limit},\ref{B-limit-muon}).

In the analysis, we have taken the magnetic field to be constant.
In fact,  this assumption is justified as long as the variation of
the magnetic field over scales of $c\Delta t$ [where $\Delta t$ is
the time segment that determines $\sigma_E$; see
Eqs.~(\ref{segments},\ref{pulsed-neutrinos})] is negligible.
Taking $E_\nu=10$ TeV and $\Gamma_{jet}=100$, we find that for the
case of pions, $c\Delta t\sim  10^{-5}~{\rm cm}\left( 10^7~{\rm
Gauss}/B\right)^{1/2}$ and for the case of  muon, $c\Delta t\sim
4\times 10^{-3}~{\rm cm} \left( 10^7~{\rm Gauss}/B\right)$. Notice
that these values of $c \Delta t$ are much smaller than the jet
size \cite{TeV} which justifies taking the magnetic field
constant.

Notice that according to Eqs.~(\ref{ratio-s},\ref{muon-ratios}),
the ratio $\sigma_E/E_\nu$ increases with energy. Thus, in the
framework of the quasi-degenerate mass scheme, for very high
energies and magnetic fields ({\it i.e.,} for $E_\nu\gg 10$~TeV
and $B\stackrel {>}{\sim} 10^9$~Gauss), the spread of the
wavepacket can overcome the wavepacket separation (see
Eq.~(\ref{as-long-as})).

{Our approximation of taking the momentum of the
particle constant over $\Delta t$ is valid only as long as $\Delta
t E_\nu\gg 1$. It is straightforward to check that for
$$(\frac{B}{10^7~{\rm Gauss}}) (\frac{E_\nu}{10~{\rm TeV}}) (\frac{100}{\Gamma_{jet}})
>10^8,$$ this condition is not fulfilled, so Eqs.~(\ref{ratio-s},\ref{muon-ratios})
 which were derived  on the basis of this approximation are not valid. For
 $$(\frac{B}{10^7~{\rm Gauss}}) (\frac{E_\nu}{10~{\rm TeV}}) (\frac{100}{\Gamma_{jet}})\sim
 10^7,$$ in both muon and pion decay
cases, $\sigma_E/E_\nu \sim 0.1$ and  the coherent correction
 to the spectrum can be significant. We will discuss this point in section
 \ref{broadening}.}

\section{Applications \label{maintain}}

\subsection{Measuring the size of the wavepakets}

{ As we discussed in the previous sections the sizes of the
neutrino wavepackets carry important information on conditions of
neutrino production: on strength of the magnetic field, on
radiation and matter density, on distance to the source. The sizes
of the packets differ for neutrinos from the muon and pion decays.
Can the sizes of the wavepackets of cosmic neutrinos be measured?
Are they relevant for observations?}

In Ref.~\cite{Stod}, it has been shown that the density matrix of
a ``stationary" beam of particles is entirely determined by the
energy spectrum and does not depend on the sizes of the
wavepackets. As a result, all the ``stationary" beams composed of
wavepackets of different sizes are equivalent and cannot be
distinguished by measurement { provided that they have the same
energy spectra}. However, in reality, even a macroscopically
stationary beam  composed of  wavepackets of finite size is not
stationary at the microscopic level. As pointed out in
\cite{Stod}, for a non-stationary beam, using  time information in
principle makes determining the size of the constituent
wavepackets a possibility.

Consider a burst of neutrinos emitted from a source at distance of
$L$ during  time interval $t_b$. If the emission is spherically
symmetric, at the time of detection neutrinos will occupy a shell
with a volume of $4\pi L^2 t_b$. The total number of emitted
neutrinos is $E_{burst}/E_\nu$, where $E_{burst}$ is the total
energy of the burst and $E_\nu$ is the average energy of
neutrinos. The average time between two successive neutrinos
passing through an area of $S$ (perpendicular to the line of sight
of the source) is \be \Delta t= \left(\frac{4\pi L^2}{S}
\right)\left(\frac{E_\nu}{E_{burst}}\right) t_b \ee $$\sim (1
~{\rm sec}) \left(\frac{L}{100~{\rm Mpc}}\right)
\left(\frac{1~{\rm m}^2}{S} \right) \left(\frac{E_\nu}{10~{\rm
TeV}}\right) \left(\frac{ 10^{52}~{\rm erg}}{E_{burst}}\right)
\left(\frac{t_b}{1~{\rm sec}}\right). $$ The time resolution of
detector can be  much better than $\Delta t = 1$ sec. As a result,
the flux will appear as non-stationary at the microscopic scale.
Suppose information on the arrival time of each wavepacket is
provided. Then, by recording the detection time with precision
better than $\Delta t$, one can in principle extract information
on the size of wavepackets composing such a flux.  If a detection
takes place at time $\tau$ from the arrival time of the ``center"
of a certain wavepacket, the wavepacket size has to be larger than
$c\tau$.  The question is whether there is any possibility to
obtain the arrival time of each wavepacket. The answer is positive
if, {\it for example}, the injection (production of pions) has
some fine structure with $\delta t<\sigma_x  \ll \Delta t$,
associated to emission  of the short $\gamma -$ burst or
gravitational waves; {\it i.e.,} in this case, the arrival time of
the ``center" (or initial moment) of the neutrino wavepacket can
be determined from the arrival time of $\gamma$'s (or
gravitational waves). Unfortunately, present models do not predict
such strict time correlation ($\delta t\ll \Delta t$) between the
$\gamma$ and $\nu$ bursts. In more realistic situations when no
information on spatial distribution of the wavepackets within the
duration of the burst is available, even recording the detection
time will not help us to determine the wavepacket size.


 The present situation argument is in line with the theorem proved in
 sec.~D.3 of \cite{Nussinov} which claims that for any flux
 composed of ``short" wavepackets with a certain { spatial}
distribution, it is possible to find an equivalent flux ({\it i.e.,} having the
 same density matrix) composed of long wavepackets with width
 $\Delta z \gg \delta z$ { provided that the long wavepackets cover  the same
part of the space in which the short wavepackets are distributed}.
Although the sizes of the wavepackets cannot be practically
determined, their effects can appear in the energy spectrum
through the coherent broadening phenomenon that is discussed in
sec.~\ref{broadening}.

\subsection{Coherence of neutrinos from various
cosmic sources \label{lost-or-not-lost}}

In this section, we check if the condition of coherence loss can
be satisfied for various sources of high energy neutrinos. The
most powerful sources of neutrinos (Gamma Ray Burst sources and
AGNs) are located at very far distances ($L>1$~Mpc). From
Eqs.~(\ref{B-limit},\ref{B-limit-muon}), { and for known values of
$\Delta m^2$} we observe that even a relatively small magnetic
field is enough to cause loss of coherence over such distances.
The magnetic fields in the GRB sources \cite{TeV} are of course
larger than these bounds. Notice that the bound  shown in
Eq.~(\ref{B-limit-muon}) is independent of the energy while the {
bound  in Eq.~(\ref{B-limit}) becomes stronger} with energy.
Taking $E_\nu$ as high as $10^{21}$~eV, from (\ref{B-limit}) we
conclude that $B$ larger than 0.1 milli-Gauss at the source is
enough to destroy the coherence of the neutrinos from the pion
decay over cosmological distances.

 Now let us discuss the cosmogenic
neutrinos. Even though the baseline for these neutrinos is of
cosmological scale, the coherence of the  neutrinos produced by
the muon decay is not lost  because in the intergalactic area,
where these neutrinos are produced, the magnetic field is too weak
to satisfy Eq.~(\ref{B-limit-muon}). However,   it is not out of
question to have
 an intergalactic magnetic field  large enough to satisfy
Eq.~(\ref{B-limit}) for $E_\nu<$ EeV, and  cause loss of coherence
for the cosmogenic neutrino from the pion decay \cite{primordial}.

For neutrinos arriving from  the center of galaxy  ($L\sim$10 kpc),
to satisfy Eq.~(\ref{B-limit}) the magnetic field at the source has
to be larger than $5\times 10^{-5}$~Gauss while to satisfy
Eq.~(\ref{B-limit-muon}) the value of the magnetic field has to be
larger than $10^5$~Gauss. If the magnetic field is between $5\times
10^{-5}$~Gauss and $10^5$~Gauss, the part of the flux from the pion
decay will lose its coherence  but the other part from the muon
decay will maintain its coherence.

 For ``solar atmospheric neutrinos'' ({\it
i.e.,} neutrinos  produced via interaction of the cosmic ray
particles with the atmosphere of the sun \cite{Seckel}), neither of
the bounds in Eq.~(\ref{B-limit}) and Eq.~(\ref{B-limit-muon}) is
satisfied: Setting $L=5 \times 10^{-6}$~pc (the distance between the
Sun and the Earth), $\Gamma_{jet}=1$, $\Delta m^2=\Delta m_{atm}^2$
and $E_\nu\sim 1$~GeV, we find that in order to satisfy
Eqs.~(\ref{B-limit}) and (\ref{B-limit-muon}) the magnetic field at
the solar surface has to be larger than $10^{5}$~Gauss  and
$10^{11}$~Gauss, respectively; that is while, the magnetic field on
the surface of the sun is less than $10^4$ Gauss so the coherence is
not destroyed.

\subsection{Coherence of
neutrinos from terrestrial sources\label{terrestrial}}

 In this section, { for comparison}  we discuss
 the coherence in
terrestrial long baseline experiments with a strong
magnetic field at the production site. Taking $E_\nu\sim 10$~GeV,
$L\sim 2 R_\oplus$, $\Gamma_{jet}=1$ and $\Delta m^2=2\times
10^{-3}~{\rm eV}^2$, from Eq.~(\ref{B-limit}) we find that the
magnetic field at the source has to be larger than $10^9$ Tesla to
obtain complete loss of coherence. For $\Delta m^2\sim 1~{\rm
eV}^2$ or larger (as suggested by LSND; see \cite{thomas} for
recent status of such models), a magnetic field of 1000~Tesla at
the source would be required to cause loss of coherence over a
baseline of the Earth size. From Eq.~(\ref{ratio-s}),  for
$E_\nu\sim 30$~GeV, $\Gamma_{jet}=1$ and $B\sim 1$ Tesla, we
obtain $\sigma_E/E_\nu\sim 10^{-6}$. In practice, the energy
resolution of the detector will be larger (worse) than this value.

For a setup such as the muon storage ring of the neutrino factory
the magnetic field is significant only at rounded edges of the
rectangular track of the muons. The neutrinos reaching the detector
are produced while the muon propagates through the side of the track
that is aligned towards the detector. The length of the wavepacket
of neutrino is therefore determined by the length of this side of
the track, $l_{str}$. Taking into account Lorentz contraction ({\it
see} sec \ref{free}) we find that the length of the wavepacket
emitted in the direction of the muon momentum is equal to \be
\sigma_x = l_{str}\frac{m_\mu^2}{E_\mu^2}. \ee Taking $E_\mu = 30$
GeV  and $l_{str} = 600$ m, we obtain $\sigma_x=0.7$~cm. From
Eq.~(\ref{ev-sep}), we find that to break the coherence for the
atmospheric mass splitting the baseline has to be larger than $\sim
0.1$ Mpc. Obviously, the coherence of the neutrinos  reaching the
detector of the neutrino factory will be maintained.

\subsection{Effects of coherent broadening
\label{broadening}}

 Coherent broadening can
lead to deformation of the energy spectrum. A well-known example
is the coherent broadening of a monochromatic line in atomic
spectroscopy. The  effects smooth down any sharp structure of
scale smaller than $\sigma_E$ in the spectrum.
If the coherent broadening is symmetric ({\it i.e.,} if the
broadening takes ``a line"   of energy $E$ to
($E-\sigma_E/2,E+\sigma_E/2$)~), the leading order correction to
the spectrum ($F$) will be given by
$$F \to \left[F+\frac{\sigma_E^2}{2}\frac{d^2F}{dE^2} \right]\frac{1}{N},$$
where $N$ is a normalization factor. However, if the broadening is
asymmetric, {\it i.e.}, if $E$ is mapped onto $(E-\sigma_E(1/2+a),
E+\sigma_E(1/2-a))$ with $a\ne 0$, the  first order correction to
the spectrum will be larger and linear in $\sigma_E$, {\it i.e.,}
$$
F \to \left(F+\frac{dF}{dE} \sigma_E a \right).
$$
In the conservative case of symmetric broadening, the effects on
spectrum can be significant if $\sigma_E\sim [(d^2
F/dE^2)/F]^{-1/2}$, where $F$ is the spectrum without taking into
account the coherent effects. Notice that the effect takes place
at production, so the distance between the source and detector
does not play  any role in this consideration. For cosmic
neutrinos, the spectrum is predicted to follow a power law
behavior: $F \propto E^{-n}$ with $n\simeq 2$. For such a
spectrum, the coherent broadening can be significant if
$\sigma_E/E_\nu\sim 0.1$.  From
Eqs.~(\ref{sigmaE-pi},\ref{sigmaE-mu},\ref{high-sigmaE}), we
observe that the coherent broadening due to scattering is
completely negligible. However, as discussed in section
\ref{magnetic}, for the decay in the magnetic field  $B\sim
10^9~{\rm Gauss}\Gamma_{jet}(10~{\rm PeV}/E_\nu)$, the ratio
$\sigma_E/E$ can be of order of 0.1 (see
Eqs.~(\ref{ratio-s},\ref{muon-ratios})). For lower values of the
magnetic field and energy, the coherent broadening is irrelevant.
As discussed in \cite{Ando}, in mildly relativistic jets with
$\Gamma_{jet}\sim 3$ present in type Ib/c supernovae, the magnetic
field  can be as high as $10^9$~Gauss. In such an environment, the
 corrections to the spectrum due to the  coherent broadening can be
significant  if $E_\nu
> {\rm few~ 10~ PeV}$. In principle, the energy loss through synchrotron radiation
can hamper production of such high energy neutrinos. Notice,
however that as shown in  \cite{Koers} through the
same mechanism that protons are accelerated in the magnetic field,
muons and charged mesons can also be accelerated. As a result,
reaching energies above PeV is possible \cite{Koers}.
Since the dependence of $\sigma_E$ on energy is non-linear (see
Eqs.~(\ref{ratio-s},\ref{muon-ratios})), the coherent effects can
change the shape of the energy spectrum.

\subsection{ Effects of incomplete averaging \label{navasaan}}

As we discussed in sec. \ref{lost-or-not-lost}, in some special
cases, the coherence of cosmic neutrinos will be maintained.
In case that coherence is not
lost, the probability of neutrino oscillation can be represented as
\be P = \bar{P} + P_{int}, \label{prob1} \ee where $\bar{P}$ is the
average probability and $P_{int}$ is the interference term which in
the two neutrino scheme can be written as
 \be P_{int} = D \cos \phi(E, L)\ . \label{inter} \ee   $\bar{P}$ and  $D$  are independent of the energy of the neutrinos,
$E$, and the baseline, $L$. The phase of oscillations in vacuum is
defined in Eq.~(\ref{phase}). In the case of three neutrino
oscillation, there will be three interference terms associated to
three different $\Delta m^2$.

   The oscillatory terms are in practice averaged to zero for the
following three reasons: (i) the integration of signal over the
energy because of the finite energy resolution of the detector; (ii)
the integration over the region of the production; and (iii) the
summation over different cosmic sources. The third reason is
specially inevitable. For example, from a single gamma burst source
at $z\sim 1$, we expect only $10^{-1}-10$ events at a 1 km$^3$ scale
detector \cite{TeV}. Fortunately, about $\sim$1000 GRB are observed
every year which can yield  at least around a few hundred neutrino
events per year at ICECUBE. When we combine the data from different
GRB sources, averaging over baselines takes place.

For typical values of the cosmic neutrino parameters, we obtain
the phase
\be
\phi =6 \times 10^{13}~ \left( \frac{\Delta m^2}{8 \times
10^{-5}{\rm eV}^2} \right) \left( \frac{L}{100~{\rm Mpc}}\right)
\left(\frac{10~{\rm TeV}}{E}\right). \ee Considering the fact that
the phases are very large, $\phi \gg 1$, we expect the oscillatory
term to average to zero. In the following, we discuss the error from
neglecting these terms.
We  discuss the procedure of averaging for two
neutrino mixing where only one interference term exists.
Generalization of our results to the three neutrino case is
straightforward.  For simplicity and definiteness, we will consider
averaging  due to integration over the neutrino energy. Integration
over source volume and summation over different sources  do not
produce new features and can be considered in a similar way.  The
``average" number of events of different types in the detector can
be written as \be \label{number} \langle N\rangle \simeq
\int_{E_{th}}dE \frac{dN}{dE}\ ,\ee where  $E_{th}$ is the detection
threshold  and
$$\frac{dN}{dE}= \frac{dN^0(E)}{dE} P(E, L).
$$
{Here} $dN^0(E)/dE$ gives the average number of events with
neutrino energy $E$ which would be observed if there was no
oscillation [{\it i.e.,} if $P(\nu_\alpha\to\nu_\beta) =
\delta_{\alpha,\beta}$].

Let us divide the whole detectable energy range into very small
intervals $\Delta E (\ll \Delta E_L)$ over which the dependence of
$dN/dE$ on energy can be neglected. The contribution of each
interval to the total number is given by a Poisson distribution
whose average and variance are both determined by $ (dN/dE) \Delta
E$.  According to the ``{\it central limit theorem"} \cite{pdg}, the
distribution of the total number is given by a Gaussian whose mean
and variance ($\sigma^2$) are both given by a sum (an integral) over
$ (dN/dE)$: $$\langle (N-\langle N \rangle)^2 \rangle=\langle
N\rangle\ ,$$ where $\langle N \rangle$ is given by
Eq.~(\ref{number}). Because of the uncertainty in the measurement of
$L$, we cannot in practice determine the value of $ \cos \phi(E,L)$
and  therefore that of $dN/dE$. However, regardless of the value of
$\phi(E,L)$, the integral of the oscillatory terms   (see
Eq.~(\ref{number})) vanishes up to a small threshold correction:
$O[\Delta E_L (dN/dE)|_{th}]\ll \sqrt{N}$. Thus, in the analysis of
cosmic neutrinos, we can safely drop the oscillatory terms.

As shown in \cite{Nussinov}, from the observational point of view,
the effect of averaging of oscillation is equivalent to the effect
of coherence loss when signals from separated wavepackets { sum up
in the detector}.  As discussed, the variance of the distribution
of the number of observed events is also the same for both cases
and is determined by the root of the number of events regardless
of the shape of the neutrino spectrum.

\subsection{Kinematical decoherence versus quantum decoherence \label{decoherence}}

Throughout this paper, we have assumed that the evolution of the
neutrino states from the source to the Earth is given by the
standard quantum mechanics. As is well-known, within the quantum
mechanics, a pure state remains  pure forever. Quantum gravity is
expected to modify the evolution of the quantum states
\cite{Hawking,Giddings}.
As long as the
modifications preserve unitarity  and the average energy and
moreover respect the second law of thermodynamics, the damping
factors appear only in front of the oscillatory terms
\cite{Fogli,Lisi}:$$P(\bar{\nu}_\alpha \to \bar{\nu}_\beta)\simeq
P(\nu_\alpha \to \nu_\beta)\simeq |U_{\alpha 1}|^2 |U_{\beta 1}|^2 +
|U_{\alpha 2}|^2 |U_{\beta 2}|^2+ |U_{\alpha 3}|^2 |U_{\beta 3}|^2
$$ $$ +2 {\rm Re}[U_{\beta 1}^* U_{\alpha 1} U_{\beta 2} U_{\alpha
2}^*] \cos \Delta_{12}e^{-L \gamma_{12}}+2 {\rm Re}[U_{\beta 1}^*
U_{\alpha 1} U_{\beta 3} U_{\alpha 3}^*] \cos \Delta_{13}e^{-L
\gamma_{13}}$$
$$+2 {\rm Re}[U_{\beta 2}^* U_{\alpha 2} U_{\beta 3} U_{\alpha 3}^*] \cos
\Delta_{32}e^{-L \gamma_{32}}\ , $$ where $\Delta_{ij}=\Delta
m_{ij}^2L/(2E_\nu)$. Thus, in the case of the cosmic neutrinos, for
which either the interference terms disappear or their effects
average to zero, the new damping factors are irrelevant. To have a
deviation from the standard picture some of the assumptions that
went into this conclusion ({\it e.g.}, conservation of the average
energy) have to be relaxed \cite{Hooper}. Our results are in accord
with \cite{tommy}.

\section{Conclusions and Summary\label{conclusion}}

In this paper, we have explored various aspects of the coherence
of the cosmic neutrinos with special emphasis on the  neutrino
flux in the (1-100) TeV energy range.
 Since  the three neutrino mass
eigenstates composing a single flavor state have different
velocities, they spatially separate on their way to the detectors
(see Eqs.
 (\ref{loss-condition},\ref{config})) provided that  the source is far enough.
Their effects in a detector cease to interfere and  the
oscillatory behavior of the probabilities  disappears. The degree
of separation depends on the size of the wavepacket, which in
turn, is determined by the properties of the environment at the
source.

We have evaluated the length of the neutrino wavepackets produced
by the pion and muon in various  environments. In an
interaction-free medium, the wavepacket sizes in the observer
frame from both the pion and muon decay are  determined  by the
lifetime of the parent particle, the energy of the parent particle
and direction of neutrino with respect to the momentum of the
decaying particle. { The size of the wavepacket  does  not depend
on the feature of the decay (two-body vs. three-body decay). In
contrast, when the parent particles undergo interactions before
decay, the wavepacket sizes  become dramatically sensitive to the
shape of spectrum. That is despite the fact that the
electromagnetic interactions of the muon and pion are practically
the same, the neutrino wavepacket sizes and their dependence on
the parameters become completely different for the pion and muon
cases.}

We have found that in an interaction-free environment,  the
wavepacket length is large  enough to maintain coherence even for
neutrinos traveling over cosmological distances. So the high
energy cosmogenic neutrinos should arrive in coherent states

Scattering of the parent charged particle
 off the particles present in the medium  decreases the
length of  the produced neutrino wavepackets. We have evaluated
the amount of shortening of the wavepacket  in a radiation
dominated environment such as the one described in \cite{TeV} as
the GRB source. We have found that in the case of neutrinos
produced by the pion, scattering can cause complete loss of
coherence (even for $\Delta m_{sol}^2$) over cosmological
distances. However, in the case of neutrinos produced by the muon
with $E_\nu\stackrel{<}{\sim}100$ TeV, the effects of scattering
are too small to cause coherence loss. In contrast, we have shown
that even a tiny magnetic field at the production region can
considerably widen the width of the neutrino wavepacket. A
magnetic field as low as a few ten Gauss at the source is enough
for  effective loss of coherence (see
Eqs.~(\ref{B-limit},\ref{B-limit-muon})). Based on the above
consideration, we have concluded that because of the sizeable
magnetic fields, all the neutrinos from remote point sources such
as GRBs and AGNs lose their coherence before reaching the Earth.
We have found that the dependence of the wavepacket size on the
magnetic field and energy is quite different for neutrinos from
the pion and muon decay. As a result, it is conceivable to have a
situation that neutrinos from pion decay lose their coherence but
those from the muon decay maintain theirs. Such a situation can be
realized for cosmogenic neutrinos as well as for ``solar
atmospheric neutrinos". For all realistic terrestrial setups the
coherence is maintained.

{ We have found that the coherent broadening can be important for
$$B{\sim} 10^9~{\rm Gauss}~ \Gamma_{jet}(10~{\rm PeV}/E_\nu).$$
Although such condition seems quite extreme, its realization is
not out of question \cite{Ando,Koers}. The dependence of
$\sigma_E$ on energy is non-linear [see
Eqs.~(\ref{ratio-s},\ref{muon-ratios})]. As a result, the coherent
effects can  change the shape of the energy spectrum of cosmic
neutrinos. }

We have studied the case in which the coherence is maintained
on the way from the source to  detector. We have found that  the
variance of the distribution of the number of events is the same as
the variance that we would expect in the absence of the oscillatory
terms. As a result, neglecting the oscillatory terms does not induce
a new source of error in determination of the neutrino parameters.

We have also  discussed the effects of a deviation of the
evolution of neutrino states from that in the standard quantum
mechanics  as one would expect in the framework of the quantum
gravity. We have shown that as long as the deviation preserves
unitarity and the average energy and moreover respects the second
law of thermodynamics, the neutrino flavor composition at the
detector is the same as the one without the deviation. To obtain a
different composition, some of these assumptions have to be relaxed.

 The neutrino pattern at the detector turns out to be the same for
 all three cases that we have studied in the present paper ({\it
 i.e.,} maintenance and loss of coherence and a deviation from
 standard quantum mechanics) and as a result, these three cases cannot be
 distinguished by observation.
Other possible applications of the results will be considered
elsewhere.

\section*{Appendix}

In the following, we study the condition for the coherence loss in
the energy-momentum space. Let us consider a neutrino field of
flavor $\alpha$ characterized by the following wavepacket
$$|\nu_\alpha \rangle=\int f(\vec{p}) |\alpha; \vec{p} \rangle d^3p, $$
where $f(\vec{p})$ is a function of the momentum which is nonzero
only in a small interval with width  $\sigma_p$. The amplitude of
finding the neutrino with flavor $\beta$, $\nu_\beta$,  at point
$\vec{x}$ and at time $t$ is
$$ \langle \nu_\beta ; \vec{x},t|\nu_\alpha \rangle=\int \sum_i
U_{\beta i}^* U_{\alpha i} e^{i[t(p+m_i^2/(2{p}))-\vec{x}\cdot
\vec{p}]} f(\vec p) d^3p \ ,
$$ where $p\equiv |\vec{p}|$. Let us denote the
 spatial resolution  by $\Delta X_D$. The
probability of finding $\nu_\beta$ in a volume of $V=(\Delta X_D)^3$
at time $t=L$ (where $L$ is the distance between the source and the
detector) is \be \label{LnuAlpha} P(\nu_\alpha \to \nu_\beta)
=\int_V \left| \langle \nu_\beta ; \vec{x},L|\nu_\alpha \rangle
\right|^2 d^3 x\ .\ee For any realistic detector, $\Delta X_D$ is
far larger than the size of the wavepacket. As a result, we can
replace the integration over $x$ with a delta function:
$$P(\nu_\alpha \to \nu_\beta) =\int \sum_{i,j} U_{\beta i}^*
U_{\alpha i} U_{\beta j} U_{\alpha j}^* e^{i(m_i^2-m_j^2)(L/2p)}
|f(\vec p)|^2 d^3 p\ . $$  $f(\vec p)$ is a relatively smooth
function over an interval of size $\sigma_p$. For $(m_j^2-m_i^2)$
satisfying the condition of loss of coherence
(\ref{loss-condition}), the integration  of the interference terms
over $\vec p$ vanishes and we therefore obtain
$$P(\nu_\alpha \to \nu_\beta)=\sum_i |U_{\beta i}|^2 |U_{\alpha
i}|^2\ .$$ Notice that this is the probability for a {\it single}
neutrino rather than the average probability for several neutrinos
states.

As discussed in \cite{Nussinov}, even if the condition
(\ref{loss-condition}) is fulfilled, for a hypothetical detector
with energy resolution, $\Delta E_D$, smaller than $\sigma_p$, the
interference terms can become relevant provided that $(\Delta
m^2L/p)(\Delta E_D/p)\stackrel{<}{\sim} 1$. Let us discuss how the
difference between the two cases can be understood within the
formalism discussed in this appendix. To calculate the total
probability of detecting $\nu_\beta$ at the detector, we first
calculated the probability of appearance of $\nu_\beta$ at a certain
point ($\vec{x}$) of the space and then integrated over $\vec{x}$.
The rationale behind this treatment was that in a realistic neutrino
detector, the neutrinos are detected through the electroweak
interactions whose range ({\it i.e.,} $m_W^{-1}\sim 10^{-16}$~cm) is
much smaller than the typical size of the wavepacket. In terms of
quantum mechanics, to a good approximation,
 the eigenstates of a realistic detection operator  are $|\nu_\beta
 ; x,t\rangle$.
In contrast, the eigenstates of a hypothetical detector with an
energy resolution of $\Delta E_D$ are states described by
$|\nu_\beta,D \rangle\equiv\int g_D(\vec p) |\nu_\beta;\vec p\rangle
d^3p $ ($\ne |\nu_\beta; \vec{x},L \rangle$), where $g_D(\vec p)$ is
a function of momentum with width $\Delta E_D$. Thus, with such a
detector, the total probability  is given by an integral over
$|\langle \nu_\beta;D|\nu_\alpha\rangle|^2$ rather than by
Eq.~(\ref{LnuAlpha}).
\section*{Acknowledgement}
We are grateful to Joern Kersten,  Georg Raffelt  and Z. Z. Xing
for useful discussions.  Y.F. would like to thank ICTP where part
of this work was done.

\end{document}